\documentclass[aps,preprint,amsmath,amssymb,superscriptaddress,nofootinbib,showkeys]{revtex4}

\usepackage{graphicx}
\begin{document}

\title{Can parallel lives provide a solution to Hardy's paradox?}

\author{\.{I}nan\c{c} \c{S}ahin}
\email[]{inancsahin@ankara.edu.tr}
 \affiliation{Department of Physics, Faculty of Sciences, Ankara University,
Ankara, Turkey}

\begin{abstract}
Parallel lives is a model which provides an interpretation of quantum theory that is both local and realistic. This model assumes that all quantum
fields are composed of point beings called "lives". Lives interact locally and have a memory of their previous interactions. The reduction of the
state vector is not included in this model: lives can be divided into different worlds. This feature resembles many worlds interpretation. However in
the parallel lives model, the division of lives into different worlds takes place locally. The parallel lives model is expected to be compatible with
special relativity, as the lives propagate at a speed that does not exceed the speed of light and interact locally. On the other hand, it is open to
paradoxes based on counterfactual propositions, as it provides a realistic interpretation of quantum theory. In this paper, we confront the parallel
lives model with the paradox proposed by Hardy \cite{Hardy:1992zz}. We show that the parallel lives model cannot overcome the dilemma in Hardy's
paradox. We discuss implications of this confrontation on special theory of relativity, and speculate a solution that we believe, fits the spirit of
the parallel lives model.

\end{abstract}



\maketitle

\section{Introduction}

Parallel lives (PL) is an ontological model that was first proposed by Brassard and Raymond-Robichaud \cite{Brassard1,Brassard2} in order to provide
a local and realistic interpretation to quantum theory (QT). The details of the PL model have been developed in Ref.\cite{Waegell:2017aqh}. According
to PL, all quantum fields are composed of point beings called "lives" moving on continuous world-lines with a speed bounded by the speed of light
\cite{Waegell:2017aqh}. Lives can only interact locally when their world-lines coincides. However, not all lives whose world-lines coincide interact
with another. Lives have a memory of their previous interactions, and this memory determines which live they will interact with. Lives that do not
interact are invisible to each other. We can say that these invisible "lives" are living in different worlds. The network of internal interactions of
a very large collection of lives forms a macroscopic system. If a live is hidden relative to one of the lives that make up the macroscopic system, it
should also be hidden relative to other lives in that macroscopic system.\footnote{Here we should note that not all lives in a macroscopic system
need to interact with each other, but they must be part of the same network of interactions. The interaction waves propagating through the
macroscopic system form a network of interactions and the memory of a distant live is shared in this way.} Thus, it is possible to have macroscopic
systems that live in parallel and hidden relative to each other. This feature recalls the many worlds interpretation \cite{Everett,DeWitt}. However,
in many worlds interpretation the entire universe split into copies, while in PL, lives locally split into relative worlds. When the state vector of
a system is reduced to one of the orthogonal terms in it, the lives that make up that system split locally into different relative worlds. Therefore,
there is no reduction of the state vector in the PL model; each orthogonal term in the superposition lives parallel in space-time. For instance,
let's consider an EPR-type experiment with two spin-1/2 particles in the singlet state, $|0,0>=\frac{1}{\sqrt
2}[|\uparrow,\downarrow>-|\downarrow,\uparrow>]$. Let {\bf A} and {\bf B} be spacelike separated macroscopic observer systems carrying Stern-Gerlach
apparatuses. After the spins become entangled in the singlet state at the midpoint between {\bf A} and {\bf B}, one moves to {\bf A} and the other to
{\bf B}. Then, the lives of spins and observers {\bf A} and {\bf B} split into relative worlds. In one world spin is up and observer measures spin-up
and in the other world spin is down and observer measures spin-down. If {\bf A$\uparrow$({\bf A$\downarrow$)}} represents observer {\bf A} measuring
spin-up (spin-down) then, the lives of {\bf A$\uparrow$} can only interact with the lives of {\bf B$\downarrow$}. Similarly, {\bf A$\downarrow$} can
interact with {\bf B$\uparrow$}, but {\bf A$\uparrow$} and {\bf A$\downarrow$} are hidden with respect to {\bf B$\uparrow$} and {\bf B$\downarrow$}
respectively. Therefore, we say that {\bf A$\uparrow$} and {\bf B$\downarrow$} are living in a world parallel to the world of {\bf A$\downarrow$} and
{\bf B$\uparrow$}.

It is often thought that Bell's theorem rules out local realistic interpretations of QT. In fact, Bell's theorem rules out local hidden variable
theories, not local realistic interpretations of QT \cite{Bell,Brassard1,Brassard2}. However, this issue is subtle and a detailed review is required.
In local hidden variable theories, the result of a measurement is given as a function of hidden variables and locally defined adjustable apparatus
parameters \cite{Bell,CHSH}. It is also assumed that experimenters have a free will to adjust apparatus parameters\footnote{Otherwise, we cannot
eliminate the superdeterminism option.} \cite{Conway-Kochen}. Let us denote the measurement result by the function $R(\lambda,a)$, where $\lambda$
and $a$ represent hidden variables and apparatus parameters respectively. The existence of the function $R(\lambda,a)$ tells us that when the values
of the parameters $\lambda$ and $a$ are given, the measurement result is uniquely determined. We will call this property {\it determinism}. In the PL
model, different possible outcomes of a measurement and observers observing these results can live in parallel in different relative worlds. Thus,
reality depends on which relative world we live in; there is no single concept of reality. Due to this multiple reality concept, some authors give up
using conventional realism \cite{Waegell2}. On the other hand, PL assumes {\it ontological reality} according to which measurement results
corresponding to orthogonal terms in the superposition {\it exist} in different relative worlds prior to measurement. This view is different from
Copenhagen interpretation, where ontological reality of the wave function is denied. PL can provide deterministic rules for the behaviors of the
lives \cite{Waegell2}. If we consider whole worlds of lives living parallel, then PL gives a deterministic model. On the other hand, each individual
observer living parallel in space-time, experiences indeterminism. For example, the observer {\bf A} performing a spin measurement (see the example
at the end of page 2) can find herself in the relative world of {\bf A$\uparrow$} or {\bf A$\downarrow$} after measurement. But she does not know in
advance which relative world she will be in. Since, the observers  cannot know in advance which one among several possible outcomes will actually
occur, the process generated by the rules of PL is completely indeterministic according to observers. Therefore, the measurement results cannot be
given as a deterministic function predicted by a local hidden variable theory. In the language of the free will theorem of Conway and Kochen
\cite{Conway-Kochen}, the response of universe to the measurement is not a function of the information accessible to the particle. The universe makes
a free decision in the neighborhood of the particle and this decision determines in which relative world the observer lives.\footnote{According to
weak anthropic principle, the observer is in one of the relative worlds just because she observes the measurement result in that relative world.}
Consequently, the locality and reality\footnote{Unless otherwise stated, reality will be used in the sense of ontological reality.} features of the
PL model do not conflict with Bell's theorem.

On the other hand, as demonstrated in several studies in the literature, the realistic interpretations of QT are inconsistent with the special theory
of relativity \cite{Hardy:1992zz,Clifton-Pagonis-Pitowsky,Pitowsky}. We should note that their arguments are based on counterfactual reasoning. When
we consider the results of actual measurements, we do not encounter paradoxes \cite{Aharonov2002}. Nevertheless, if we have a realistic model where
the wave function or say the probability distribution of the possible outcomes exists prior to the measurement then counterfactual propositions
become somewhat legitimate \cite{Vaidman}. Therefore, any model that claims to provide a realistic interpretation of QT must be confronted with
counterfactual paradoxes. In this context, we confront PL model with the second paradox in Hardy's paper \cite{Hardy:1992zz}. As we will see, in PL
some counterfactual propositions become part of the reality in various alternative worlds. This has interesting implications for the theory of
relativity, which we will examine.

\section{Revisiting Hardy's paradox}

\begin{figure}
\includegraphics[scale=2]{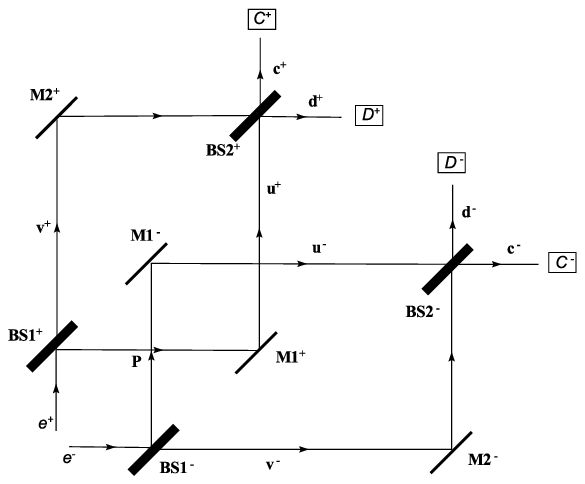}
\caption{Scheme of Hardy's gedankenexperiment \cite{Hardy:1992zz}. $BS1^+, BS1^-, BS2^+, BS2^-$ represent beam splitters and $M1^+, M1^-, M2^+, M2^-$
represent mirrors. $C^+, D^+, C^-, D^-$ are detectors. \label{fig1}}
\end{figure}

In 1992 Hardy \cite{Hardy:1992zz} proposed a gedankenexperiment consists of two Mach-Zehnder interferometers, one for positrons and one for electrons
Fig.\ref{fig1}. The experiment is designed so that $u^+$ and $u^-$ paths of these two Mach-Zehnder interferometers overlap. If the positron and
electron take $u^+$ and $u^-$ paths then they will meet at $P$ and annihilate one another. Pair annihilation is expressed in Hardy's notation as
\begin{eqnarray}
\label{pair annihilation} |u^+> |u^-> \to |\gamma>.
\end{eqnarray}
Using the experimental setup shown in the Fig.\ref{fig1}, Hardy first demonstrated an inequality-free version of the Bell's theorem. Hardy secondly
demonstrated that if the "elements of reality" corresponding to Lorentz-invariant observables are themselves Lorentz invariant, then realistic
interpretations of quantum mechanics are incompatible with special theory of relativity. For the purpose of this paper we will concentrate on his
second result. The summary of the reasoning that led him to this conclusion is as follows: Consider three different reference frames: $LAB$, $S^+$
and $S^-$ frames of reference. In $LAB$ frame, the measurements on electron and positron are simultaneous. The relative velocities of $S^+$ and $S^-$
frames to $LAB$ frame are so arranged that these measurements are not simultaneous with respect to $S^+$ and $S^-$. According to $S^+$ frame the
measurement on the positron occurs before the electron arrives at $BS2^-$ and according to $S^-$ frame the measurement on the electron occurs before
the positron arrives at $BS2^+$. Let's denote the initial electron-positron states by $|e^-> |e^+>$. After the particles pass point P, but before
they reach $BS2^{\pm}$ the initial state evolves to
\begin{eqnarray}
\label{*}
|e^-> |e^+>\to \frac{1}{2}(-|\gamma>+i|u^+> |v^->+i|v^+> |u^->+|v^+> |v^->).
\end{eqnarray}
Since this state is orthogonal to $|u^+> |u^->$, according to an observer in the $LAB$ frame positron and electron cannot take $u^+$ and $u^-$ paths
simultaneously. The beam splitters $BS2^{\pm}$ perform the following transformations:
\begin{eqnarray}
\label{BS2} |u^\pm> \to \frac{1}{\sqrt{2}}(i|d^{\pm}>+|c^{\pm}>),\;\;\;\;|v^\pm> \to\frac{1}{\sqrt{2}}(i|c^{\pm}>+|d^{\pm}>).
\end{eqnarray}
Using equations (\ref{*}) and and (\ref{BS2}) we see that the state vector reduces to $|d^+> |d^->$ with a probability of $\frac{1}{16}$. Hence, in
$\frac{1}{16}$th of the experiments both $D^+$ and $D^-$ detectors receive signals.

Now, let's examine the same experiment according to observers in the $S^-$ and $S^+$ frames. According to $S^-$ when the electron passes through
$BS2^-$ but the positron has not yet reached $BS2^+$, the following state is obtained:
\begin{eqnarray}
\label{*S-}  \frac{1}{2}(-|\gamma>-\frac{1}{\sqrt{2}}|u^+> |c^->+\frac{i}{\sqrt{2}}|u^+> |d^->+i\sqrt{2}|v^+> |c^->).
\end{eqnarray}
Here we use (\ref{*}) and transformations for $|u^->$ and $|v^->$ in (\ref{BS2}). When the electron is detected in $D^-$, then the state vector is
reduced to
\begin{eqnarray}
\label{**} |u^+> |d^->.
\end{eqnarray}
Then, the observer in the $S^-$ frame infers that positron takes $u^+$ path. On the other hand, according to $S^+$ when the positron passes through
$BS2^+$ but the electron has not yet reached $BS2^-$, the following state is obtained:
\begin{eqnarray}
\label{*S+}  \frac{1}{2}(-|\gamma>-\frac{1}{\sqrt{2}}|c^+> |u^->+\frac{i}{\sqrt{2}}|d^+> |u^->+i\sqrt{2}|c^+> |v^->).
\end{eqnarray}
Here we use (\ref{*}) and transformations for $|u^+>$ and $|v^+>$ in (\ref{BS2}). When the positron is detected in $D^+$, then the state vector is
reduced to
\begin{eqnarray}
\label{***} |d^+> |u^->.
\end{eqnarray}
Then, the observer in the $S^+$ frame infers that electron takes $u^-$ path.\footnote{Here we should note that the inferences of observers in $LAB$,
$S^+$ and $S^-$ frames about particle trajectories ($u^+$ and $u^-$) are counterfactual. They don't make measurements to determine real paths, but
they infer these results from $D^+$ and $D^-$ detections via counterfactual reasoning.}

Hardy used EPR's \cite{EPR} "element of reality" criterion. If a system is in an eigenstate of an operator corresponding to an observable, then we
can predict certainly the result of the measurement of this observable. Therefore, according to EPR's reality criterion, the value of this observable
(which is the eigenvalue of the observable corresponding to the system eigenstate) is an element of reality even if the measurement is not performed.
We can define the operators $U^{\pm}=|u^{\pm}><u^{\pm}|$. Since the vectors $|u^+>$ and $|u^->$ are eigenvectors of $U^{\pm}$, there exist elements
of reality associated with paths $u^+$ and $u^-$. However, as we have shown, the reference frames $S^+$ and $S^-$ infer that electron and positron
take the paths $u^-$ and $u^+$ respectively. If the elements of reality corresponding to Lorentz-invariant observables are themselves Lorentz
invariant, then these inferences must be true for all inertial frames. On the contrary, as shown previously it is not true for the $LAB$ frame. To
summarize very briefly, what Hardy did is that he associated counterfactuals about particle paths with elements of reality. Then, he showed that
elements of reality corresponding to these paths are not Lorentz invariant.

\begin{figure}
\includegraphics[scale=2]{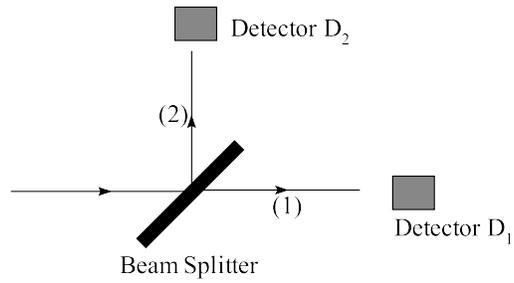}
\caption{Single photon on a beam splitter. \label{fig2}}
\end{figure}

As stated in his article, Hardy's result can be applied to any realistic interpretation of QT which assumes that particles have real trajectories. In
PL model, lives move on real trajectories in space-time. Therefore, confrontation of the PL model with Hardy's paradox can have important
consequences. Before examining Hardy's paradox in the PL model, let's examine lives of a single photon on a beam splitter and in a Mach-Zehnder
interferometer. In Fig.\ref{fig2} we show a single photon on a 50-50 beam splitter. Incident photon can either be transmitted along path (1) or
reflected along path (2). Each path has 50\% probability. Assume that an observer performs a measurement using photon detectors to determine the path
along which the photon moves. This measurement causes an entanglement between photon paths and measurement apparatus:
\begin{eqnarray}
\label{measurement1} |\psi>=\frac{1}{\sqrt {2}}|1_{\gamma}> |1_m>+\frac{1}{\sqrt {2}}|2_{\gamma}> |2_m>
\end{eqnarray}
where, $|1_{\gamma}>$ represents the photon state in path (1) and $|1_m>$ represents the state of measurement apparatus measuring a photon in path
(1). Similar definitions hold for $|2_{\gamma}>$ and $|2_m>$. Furthermore, we can also say that the observer is entangled with photon paths. By
looking at the result of the measurement, the observer can decide to behave in one way or another. For instance, assume that if the photon takes path
(1), then the observer will have lunch. On the other hand, if the photon takes path (2), then she will be on diet. Thus, we can write
\begin{eqnarray}
\label{measurement2} |\psi>=\frac{1}{\sqrt {2}}|1_{\gamma}> |1_\text{o}>+\frac{1}{\sqrt {2}}|2_{\gamma}> |2_\text{o}>.
\end{eqnarray}
Here, subscript "o" denotes the observer. The description of the experiment within the PL model can be given as follows: The lives of the incident
photon are divided into two group of lives living in the same world. One of them takes path (1) and the other takes path (2). When the lives of the
photons moving on paths (1) and (2) meet with the detectors, lives of each detector, subsequently lives of the measurement apparatus and the observer
are divided into two different worlds. In one world $D_1$ detects a signal but $D_2$ does not detect any signal, in another world $D_1$ does not
detect any signal but $D_2$ detects a signal. Consequently, in one world observer measures a photon moving on path (1) and in the other world she
measures a photon moving on path (2). These two worlds are hidden with respect to each other.

\begin{figure}
\includegraphics[scale=2]{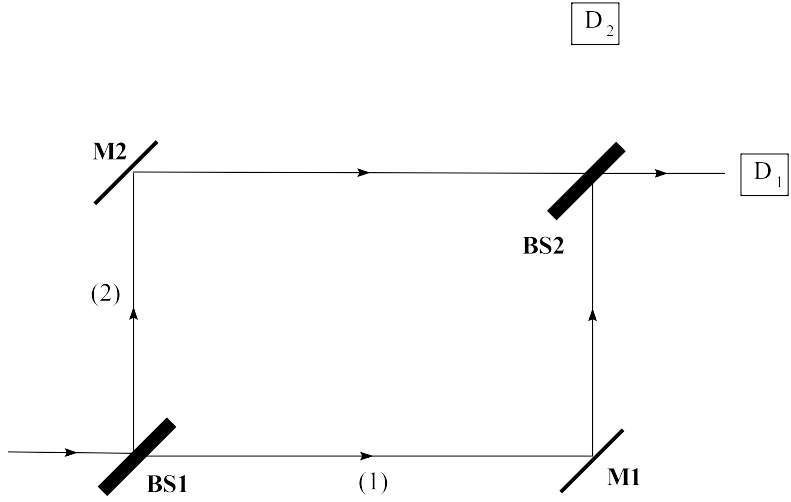}
\caption{Single photon in a Mach-Zehnder interferometer. \label{fig3}}
\end{figure}

Now, let's consider a single photon in a Mach-Zehnder
interferometer Fig.\ref{fig3}. Due to destructive interference, $D_2$ detector does not detect any signal. Therefore, in this case photon paths are not
entangled with the measurement apparatus or the observer. Hence, the lives of the measurement apparatus and the observer are not divided into relative worlds.
When the initial photon passes through the first beam splitter, its lives are divided into two group of lives, one going through the path (1) and the other
going through the path (2). These two group of lives moving on paths (1) and (2), exist in the same world. In the second beam splitter they interact with
each other and produce the usual interference effects.

\begin{figure}
\includegraphics[scale=2]{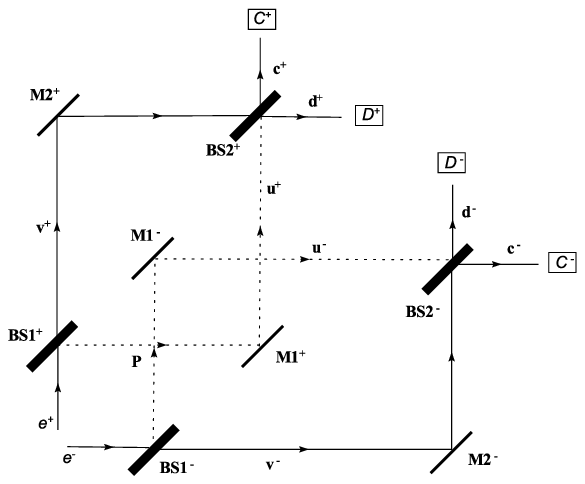}
\caption{Diagram representing the lives observed in the $LAB$ frame. Dotted lines represent hidden lives living in parallel.\label{fig4}}
\end{figure}

Finally, let's try to examine Hardy's paradox in the framework of the PL model. In the $LAB$ frame of reference, both of the particles reach second
beam splitters simultaneously. Just before they reach second beam splitters the state of the system is given by (\ref{*}). This state is orthogonal
to $|u^+> |u^->$. Therefore, according to an observer in the $LAB$ frame positron and electron cannot take $u^+$ and $u^-$ paths together.
Accordingly, lives of the positron and electron moving on paths $u^+$ and $u^-$ must be hidden in the world of the $LAB$ frame, i.e. they are living
parallel to the $LAB$ frame.\footnote{This is evident from equation (\ref{*}), but it is also conceivable from pair annihilation process at point
$P$. If the particles take paths $u^+$ and $u^-$, then pair annihilation occur. In this case, the positron and electron turn into two photon and do
not leave any signal in the detectors $D^+,D^-,C^+,C^-$. If we have additional photon detectors, we can capture photon signals from pair
annihilation. However, since we restrict ourselves to the situation where both $D^+$ and $D^-$ detectors detect signals, there should be no pair
annihilation in the world of the $LAB$ frame.} In Fig.\ref{fig4} we show a diagram representing the lives observed in the $LAB$ frame.

\begin{figure}
\includegraphics[scale=2]{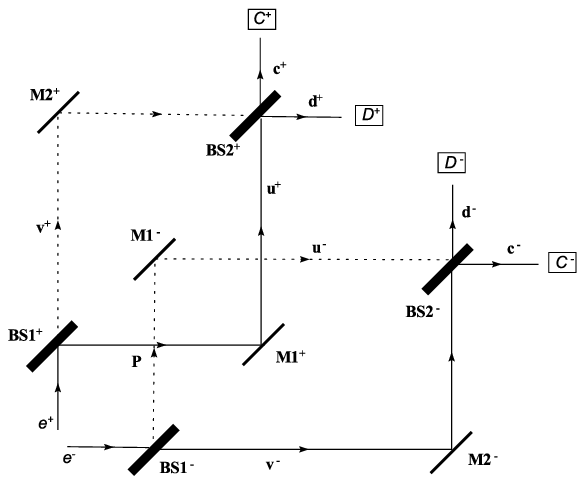}
\caption{Diagram representing the lives observed in the $S^-$ frame. Dotted lines represent hidden lives living in parallel. \label{fig5}}
\end{figure}

\begin{figure}
\includegraphics[scale=2]{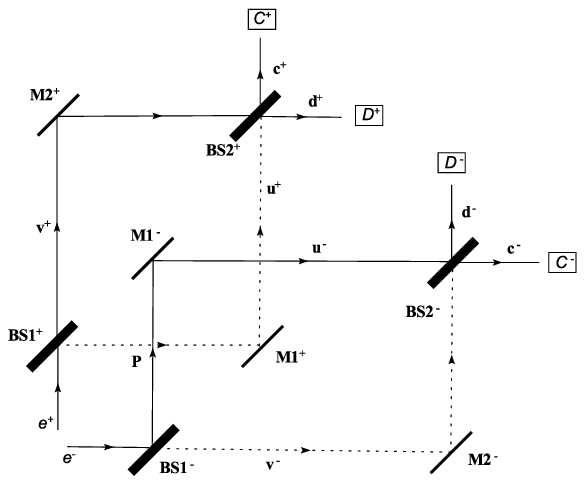}
\caption{Diagram representing the lives observed in the $S^+$ frame. Dotted lines represent hidden lives living in parallel. \label{fig6}}
\end{figure}

Let's depict the same experiment according to an observer in $S^-$ frame of reference. Due to the relativity of simultaneity, the positron has not
yet reached $BS2^+$ as soon as the electron passes through $BS2^-$. At this instant, the system is described by the state given in (\ref{*S-}).
Within a very short time, electron can reach $C^-$ and $D^-$. Hence, the following entangled state is obtained:
\begin{eqnarray}
\label{*S-entangled}  &&-\frac{1}{2}|\gamma> |C^-=0;D^-=0> -\frac{1}{2\sqrt{2}}|u^+> |c^-> |C^-=1;D^-=0> \nonumber \\
&&+\frac{i}{2\sqrt{2}}|u^+> |d^-> |C^-=0;D^-=1> +\frac{i}{\sqrt{2}}|v^+> |c^-> |C^-=1;D^-=0>
\end{eqnarray}
where, $|C^-=0,1;D^-=0,1>$ is the state of the measurement apparatus; 1 represents detection of a particle and 0 represents a null value (no
detection). Consequently, lives of the observer and experimental apparatus split into four different worlds, corresponding to orthogonal terms in the
superposition (\ref{*S-entangled}). Since we restrict ourselves to the situation where $D$ detectors detect signal, we consider the relative world of
$S^-$ described by the third term in (\ref{*S-entangled}). In this relative world, lives of the positron take $u^+$ path and lives moving on paths
$u^-$ and $v^+$ are hidden. In Fig.\ref{fig5} we show the lives of the experimental apparatus observed in the $S^-$ frame. On the other hand,
according to an observer in $S^+$ frame of reference, the electron has not yet reached $BS2^-$ as soon as the positron passes through $BS2^+$. At
this instant, the system is described by the state given in (\ref{*S+}). Within a very short time, positron can reach $C^+$ and $D^+$. Hence, the
following entangled state is obtained:
\begin{eqnarray}
\label{*S+entangled}  &&-\frac{1}{2}|\gamma> |C^+=0;D^+=0> -\frac{1}{2\sqrt{2}} |u^-> |c^+> |C^+=1;D^+=0> \nonumber \\
 &&+\frac{i}{2\sqrt{2}} |u^-> |d^+> |C^+=0;D^+=1> +\frac{i}{\sqrt{2}} |v^-> |c^+> |C^+=1;D^+=0>.
\end{eqnarray}
In the relative world of $S^+$ described by the third term in (\ref{*S+entangled}), lives of the electron take $u^-$ path and lives moving on paths
$u^+$ and $v^-$ are hidden. The lives of the experimental apparatus observed in the $S^+$ frame is given in Fig.\ref{fig6}.

To summarize, the lives of particles in the worlds of different reference frames are different from each other. The lives moving on path $u^+$ are
part of the world of the $S^-$ frame, but not part of the worlds of the $S^+$ and $LAB$ frames. Similarly, the lives moving on path $u^-$ are part of
the world of the $S^+$ frame, but not part of the worlds of the $S^-$ and $LAB$ frames. However, we should note that actually the lives were there
all along. The only thing that changes from one frame of reference to another is whether lives of the particles interact or not with the apparatus.
As we have discussed in the introduction, noninteracting lives are hidden, and the observer cannot experience them in her world.

The fact that different reference frames live parallel to each other in different worlds seems to fit the logic of PL at first sight. However as we
will see, there is a problem we have to overcome. The observer in each reference frame observes not only the experimental apparatus but also the
observer in the other reference frame. For instance, let's denote the lives of the observer in the $S^-$ frame of reference observing the measurement
results $C^-=0$ and $D^-=1$ by $\bf{O_{S^-}(D^-=1)}$. Denote also the lives of the experimental apparatus with $C^-=0$ and $D^-=1$ by
$\bf{A(D^-=1)}$. When these two lives meet, they merge to form a bigger set of lives that we will denote as
\begin{eqnarray}
\label{S^-&A} \bf{O_{S^-}(D^-=1)}\oplus\bf{A_{S^-}(D^-=1)}.
\end{eqnarray}
Here, the subscript $S^-$ in $\bf A$ represents the configuration of the lives of the apparatus observed by $O_{S^-}$ (configuration in
Fig.\ref{fig5}). Let the lives $\bf{O_{S^-}(D^-=1)}$ and $\bf{A_{S^-}(D^-=1)}$ meet the lives of the observer in $S^+$ frame before positron reaches
$BS2^+$, then the following set of lives is obtained:
\begin{eqnarray}
\label{S^-&A$S^+} \bf{O_{S^-}(D^-=1)}\oplus\bf{A_{S^-}(D^-=1)}\oplus\bf{O_{S^+}(D^-=1)_{3.}}
\end{eqnarray}
where the subscript "3." indicates that this describes a "third-person perspective": observer in $S^-$ frame observes in her world another "observer"
in the $S^+$ frame of reference which she denotes $\bf{(O_{S^+}})_{3.}$.\footnote{We borrow this terminology from Ref.\cite{Mueller:2017cdn}.
However, Ref.\cite{Mueller:2017cdn} used this terminology in the context of algorithmic information theory and did not apply it to relativistic
observers.} After a while, positron also passes $BS2^+$ and is then detected. The detection of the positron causes the lives of the apparatus and the
observers split into relative worlds: In one world we obtain $C^+=0$, $D^+=1$ and in the other world $C^+=1$, $D^+=0$. Since we consider
$D^-=1,D^+=1$ case, lives of the joint system become
\begin{eqnarray}
\label{S^-&A$S^+*} \bf{O_{S^-}(D^-=1;D^+=1)}\oplus\bf{A_{S^-}(D^-=1;D^+=1)}\oplus\bf{O_{S^+}(D^-=1;D^+=1)_{3.}}.
\end{eqnarray}
The above expression reflects first-person perspective of the observer $O_{S^-}$.\footnote{we omit the subscript "1." for abbreviation.} In this
perspective $D^-=1$ and $D^+=1$ detections occurred due to lives coming from  $v^-$ and $u^+$  paths (see Fig.\ref{fig5}). Therefore, lives moving on
paths $v^-$ and $u^+$ are part of the history of (\ref{S^-&A$S^+*}). On the other hand, first-person perspective of the observer $O_{S^+}$ has
experienced an other history. According to $O_{S^+}$, $D^-=1$ and $D^+=1$ detections occurred due to lives coming from  $u^-$ and $v^+$  paths (see
Fig.\ref{fig6}). In the first-person perspective of the observer $O_{S^+}$, we can write the following world of lives:
\begin{eqnarray}
\label{S^+&A$S^-*} \bf{O_{S^-}(D^-=1;D^+=1)_{3.}}\oplus\bf{A_{S^+}(D^-=1;D^+=1)}\oplus\bf{O_{S^+}(D^-=1;D^+=1)}.
\end{eqnarray}
From the analysis we performed above, we get the following odd-looking result: first-person and third-person perspectives of the same observer belong
to different worlds. The observer $\bf{(O_{S^+}})_{3.}$ in the world of $\bf{(O_{S^-}})_{1.}$ lives parallel to the world of $\bf{(O_{S^+}})_{1.}$.
But if quantum laws apply equally to all observers, then $\bf{(O_{S^+}})_{3.}$ should not observe that the positron is detected before the
electron.\footnote{Otherwise the state vector is reduced to (\ref{***}), which indicates that electron takes $u^-$ path. However, this is erroneous
as seen from (\ref{S^-&A$S^+*}).} However, this result is incompatible with the relativity of simultaneity: $\bf{(O_{S^+}})_{3.}$ is moving relative
to $\bf{(O_{S^+}})_{1.}$, and the time order of the detection events should be reversed. Consequently, we encounter a discrepancy between special
relativity and the PL model.

Nevertheless, we need to say that such a discrepancy does not arise for any interpretation of QT that does not accept the reality of anything other
than the measurement outcomes. According to such an interpretation, the paths $u^-$, $u^+$, $v^-$ and $v^+$ are just mathematical auxiliary concepts;
they are not related to reality.

\section{Speculations on theory of relativity}

If we persist in the realistic interpretations of QT, the discrepancy with the theory of relativity needs to be resolved. One solution to this
discrepancy is to modify the theory of relativity by proposing a preferred frame of reference. Such a modification of the theory has been discussed
for a long time \cite{CliffordWill}. However, there are obscurities in this approach, such as which criteria should be used to determine the
preferred frame of reference.\footnote{One possible candidate for preferred frame of reference is the frame in which the cosmic microwave background
is isotropic \cite{CliffordWill,Coleman:1997xq}. However, there is not any apparent reason why this frame should be the preferred frame of
reference.} In this paper we will make the following speculation which we believe offers a solution to the discrepancy and also fits the spirit of
the PL model: {\it There is no particular preferred frame of reference, but for each frame there is always a world in which that frame is preferred.
The world observed from an observer's first-person perspective is the world where the observer's stationary frame is preferred. Lorentz
transformations\footnote{Conventional Lorentz transformations in the symmetrical form.} are defined between first-person perspectives of observers on
different inertial frames of reference.}

According to the assumptions above, lives of each observer split into infinitely many worlds; one of them corresponds to observer's first-person
perspective and others correspond to third-person perspectives of some other observers. Suppose that $S_1$,$S_2$,...,$S_n$ are different inertial
reference frames. Then lives of the observer of each reference frame ${S_i}$, $i\in \{1,2,...n\}$ split into n relative worlds. One of them is the
world observed in the first-person perspective of the observer in the frame ${S_i}$. In this world we denote the lives of the observer in ${S_i}$ by
$\bf{(O_{S_i}})_{1.}$. All other observers are in the third-person perspective and denoted by $\bf{(O_{S_1}})_{3.}$,
$\bf{(O_{S_2}})_{3.}$,..,$\bf{(O_{S_{i-1}}})_{3.}$, $\bf{(O_{S_{i+1}}})_{3.}$,..,$\bf{(O_{S_{n}}})_{3.}$. As is known, Lorentz transformations have a
symmetrical form, i.e. the transformations $S_i\to S_j$ and $S_j\to S_i$, $(i\neq j)$ have exactly the same form, up to the sign in front of the
velocity. This feature implies that we cannot distinguish one frame of reference from another. In our assumptions, a Lorentz transformation from
$S_i$ to $S_j$, essentially defines a transformation from $\bf{(O_{S_i}})_{1.}$ to $\bf{(O_{S_j}})_{1.}$. $\bf{(O_{S_i}})_{1.}$ and
$\bf{(O_{S_j}})_{1.}$ live parallel in different worlds and each is the preferred observer in her own world. We interpret the symmetry feature of
Lorentz transformations as the equivalence of the worlds of $\bf{(O_{S_i}})_{1.}$ and $\bf{(O_{S_j}})_{1.}$ in defining the laws of nature.

One can then ask the transformations between observers in the first-person and third-person perspectives, i.e. transformations
$\bf{(O_{S_i}})_{1.}\to\bf{(O_{S_j}})_{3.}$? In this case $S_i$ is the preferred frame of reference. Therefore, the order of events observed by
$\bf{(O_{S_i}})_{1.}$ determine the physical behavior in Hardy's gedankenexperiment. For instance, if $S_i$ coincides with $S^{-}$  frame of
reference then the detection of the electron takes place before the detection of the positron and hence, lives of the joint system of observers and
the apparatus is given by (\ref{S^-&A$S^+*}). All other observers in the world of $\bf{(O_{S_i}})_{1.}$ should observe same order of detection
events. Therefore $\bf{(O_{S_j}})_{3.}$ observes variable speed of light, and hence the transformations $\bf{(O_{S_i}})_{1.} \leftrightarrows
\bf{(O_{S_j}})_{3.}$ does not obey conventional Lorentz transformation formula. To be precise, assume that the detections in the $D^+$ and $D^-$
detectors are synchronized with light pulses from outer point $K$. According to $\bf{(O_{S_i}})_{1.}$, these light pulses propagate with a speed $c$.
Then, according to $\bf{(O_{S_j}})_{3.}$, $(i\neq j)$ speeds of these light pulses moving from $K$ to $D^+$ and $D^-$ can vary and their values may
no longer be $c$. The discussion of what the explicit forms of these transformations is beyond the purpose of this paper. However, we would like to
draw attention to the following point: Whatever new transformations are, it may not be valid globally. For instance, speed of light from emission
event at $K$ to the absorption event at $D^-$ may not be equal to the speed of light moving between other two events.\footnote{Even the speeds of
light pulses from $K$ to $D^-$ and $K$ to $D^+$ may not be equal.} Therefore, the transformation used, varies depending on which events it is used
for. This gives us locally defined transformations. This peculiar situation becomes understandable to some extent if we realize that the world of
$\bf{(O_{S_i}})_{1.}$ emerge as a result of the entanglement of the Hardy's experimental setup with $\bf{(O_{S_i}})_{1.}$. Accordingly, in this world
we can attribute a special meaning to the signal events in $D^-$ and $D^+$ detectors. We can consider some kind of transformation which gives
conventional Lorentz transformation formula for events not associated with Hardy's experimental setup, but gives a new or modified transformation
formula for signal events in the $D^-$ and $D^+$ detectors. Obviously, this new transformation violates the Lorentz symmetry. However, the Lorentz
symmetry is violated only for events associated with quantum entanglement between the observer and some quantum system. Therefore, we can say that
Lorentz symmetry is {\it almost} valid.

As it was said by Barbour \cite{Barbour}, Einstein did not create a theory of clocks and duration from first principles. He avoided ever having to
address the physical working of rods and clocks; they were always treated separately as independent entities in both relativity theories. Their
properties were not deduced from the inner structure of the theory, but were simply required to accord with the relativity principle \cite{Barbour}.
We claim that QT gives actual physical working of rods and clocks. But we should be open to the idea that the relativity principle may not be
absolute, and can be violated for certain events associated with quantum entanglement.

Finally, we want to discuss how we should interpret the non-equivalence of an observer's first-person and third-person perspectives. What exactly
does this mean? Does this mean that the observer $\bf{(O_{S_j}})_{3.}$ in $\bf{(O_{S_i}})_{1.}$'s world is an unconscious being, such as a zombie or
a robot? This is not what we intend to say. If we want to explain with the example of Hardy's gedankenexperiment we discussed in the previous
section, we can say that the measurement performed by $\bf{(O_{S_i}})_{1.}$ and her conscious perception causes the state vector to
collapse.\footnote{Of course, there is no state vector collapse in the PL model. But since we think many physicists are more familiar with this
terminology, we use the term "collapse" for clarity.} But this does not mean that $\bf{(O_{S_j}})_{3.}$ is an unconscious being. It simply means that
in $\bf{(O_{S_i}})_{1.}$'s world, $\bf{(O_{S_j}})_{3.}$'s perception of the measurement result has no effect on the state vector's collapse; all
observers in different reference frames respect the order of events and recorded history that the observer $\bf{(O_{S_i}})_{1.}$ sees on Hardy's
experimental setup. On the other hand, if we repeat or perform another experiment, lives will split again and $\bf{(O_{S_j}})_{3.}$ can find herself
in the world of her first-person perspective where her frame of reference is the preferred frame. As soon as this happens, the subscript "3." should
be replaced by "1.".

\section{Conclusions}

PL is a model that is expected to be compatible with the relativity theory because it includes the local interactions of lives and their motions that
do not exceed the speed of light. However, we negated this expectation by showing that the PL model could not overcome the paradox suggested by
Hardy. Our results can also be applied to many world interpretation where counterfactual propositions assumed to be part of reality in different
alternative worlds, or any realistic interpretation of QT that assumes real particle trajectories. But we want to emphasize that there is no conflict
between the special theory of relativity and QT for approaches and interpretations that regard state vectors as auxiliary mathematical concepts and
do not relate them to reality. Therefore, one way to overcome the Hardy's paradox is to adopt such an approach. On the other hand, if we insist on a
realistic interpretation as we have just mentioned, we must accept the possibility that Lorentz symmetry is violated. Such a Lorentz symmetry
violation can be realized by choosing a preferred frame of reference, as noted in Hardy's original paper \cite{Hardy:1992zz}. In section III, we have
made an interesting speculation which we believe offers a solution to the discrepancy between QT and special theory of relativity, and also fits the
spirit of the PL model.

\newpage

\end{document}